\documentclass[12pt,preprint]{aastex}

\newcommand{\etal}{et al.}

\shorttitle{Two Populations of AGN}
\shortauthors{Draper \&  Ballantyne}

\begin{document}

\title{A Tale of Two Populations: The Contribution of Merger and Secular Processes to the Evolution of Active Galactic Nuclei}


\author{A. R. Draper and D. R. Ballantyne}
\affil{Center for Relativistic Astrophysics, School of Physics,
  Georgia Institute of Technology, Atlanta, GA 30332}
\email{aden.draper@physics.gatech.edu}

\begin{abstract}
Due to the co-evolution of supermassive black holes and their host galaxies, understanding the mechanisms that trigger active 
galactic nuclei (AGN) are imperative to understanding galaxy evolution and the formation of massive galaxies.   
It is observationally difficult to determine the trigger of a given AGN due to the difference between the AGN lifetime and triggering timescales. 
Here, we utilize AGN population synthesis modeling to determine the importance of different AGN triggering mechanisms.
An AGN population model is computed by 
combining an observationally motivated AGN triggering rate and a theoretical AGN light curve.  The free parameters of the AGN light 
curve are constrained by minimizing a $\chi^2$ test with respect to the observed AGN hard X-ray luminosity function.  The observed 
black hole space density, AGN number counts, and X-ray background spectrum are also considered as observational constraints.  
It is found that major mergers are not able to account for the entire AGN population.  Therefore, non-merger processes, such as 
secular mechanisms, must also trigger AGN.  Indeed, non-merger processes are the dominant AGN triggering mechanism at $z$ $\lesssim$ 1--1.5.  
Furthermore, the shape and evolution of the black hole mass function of AGN triggered by major mergers is 
intrinsically different from the shape and evolution of the black hole mass function of AGN triggered by secular processes.  

\end{abstract}


\section{Introduction}
\label{sect:intro}

In less than a decade it was determined that not only do all massive galaxies harbor a supermassive black hole at 
their center \citep{KR95}, but also that the evolution of the galaxy is intrinsically linked to the growth of the central black hole 
\citep[e.g.,][]{M98,T02}.  However, the mechanism responsible for the co-evolution of a supermassive black hole and its host galaxy is 
still under investigation.  As accreting supermassive black holes, generally referred to as active galactic nuclei (AGN), 
can, over their lifetime, radiate an amount of energy comparable to the binding energy of their host galaxy \citep[e.g.,][]{SR98,F99,A10}, 
it is believed that this AGN feedback will 
affect star formation in the host galaxy \citep[e.g.,][]{F99, H06, Lag08, Tri09, B10, H11, HS11}.  Galaxy wide processes can 
also affect the central supermassive black hole.  For example, theoretical and observational evidence suggests that AGN can be triggered by major mergers 
of massive gas rich galaxies \citep[e.g.,][]{S88, H89, C90, KH00, H06} or by secular processes not connected to major mergers, such as supernova winds, 
stellar bars, cold-flow accretion, interactions with other massive galaxies or satellite galaxies, or minor mergers \citep[e.g.,][]{C03, KK04, V05, D07, P07, C09, KJ10, Bo11, O11}.

Simulations show that mergers of gas rich galaxies cause gas and dust to lose angular momentum and fall into the central 
regions of the galaxy \citep[e.g.,][]{BH96}.  The resulting nuclear gas reservoir will be consumed by star formation in the host bulge and 
accretion flows onto the resident supermassive black hole \citep{H06, H06b}.  According to the galaxy evolution model explored by \citet{H09}, AGN 
feedback will then evolve the AGN host galaxy across the color-magnitude diagram (CMD) from the massive end of the blue cloud, through the 
green valley, and onto the red sequence (but see Schawinski et al. 2009 and Cardamone et al. 2010).  
Secular processes, such as supernova explosions, stellar bars, minor mergers and  
interactions, will also release angular momentum from gas and dust leading to a similar reservoir of gas in the 
central region of the galaxy \citep[e.g.,][]{C03,D07,C09, O11}.  Despite mergers and secular mechanisms both leading to the accumulation of gas 
and dust deep in the galactic potential well, mergers and secular processes have very different galaxy wide effects 
and roles in galaxy evolution.  Major mergers are violent processes which, as with the Antennae galaxies, can destroy 
galactic structure, causing the merger remnant to relax into a bulge-dominated system \citep[e.g.,][]{HH06}.  
Secular evolution, however, is 
not likely to disturb the morphology or large-scale structure of the host galaxy \citep[see][]{G09}.  Despite the systemic differences between 
galaxy evolution and AGN activity due to major mergers and secular processes, it is difficult to observationally determine 
which mechanism is responsible for a given AGN; tidal tails are often faint, bulge-dominated systems can be re-triggered by secular processes, 
and the timescales for AGN activity tend to be longer than the timescales for a merger remnant galaxy to dynamically relax \citep{S10a}.  

Despite the observational difficulties in determining how a particular AGN has been triggered, there is observational evidence 
that major mergers are not the dominant AGN triggering mechanism, at least not at $z$ $\lesssim$ 2.  By determining the S\'ersic indices 
of massive galaxies at $z$ = 2--3, \citet{Wein11} found that $\sim$65$\%$ of AGN hosts at this redshift range have S\'ersic indices 
indicative of disky morphologies.  Observations also show that in the redshift ranges $z$ = 1.5--3 \citep{S11} and $z$ = 1.5--2.5 \citep{K12}, the majority of moderate luminosity AGN are hosted by disk galaxies.  By considering galaxy pairs at $z$ 
$\lesssim$ 2, \citet{W11} find that the wet major merger rate is too low to account for the majority of AGN activity at $z$ $\lesssim$ 2.
When investigating AGN in zCOSMOS, \citet{Sil11} found that $\sim$20$\%$ of moderate luminosity AGN at 0.25 $<$ $z$ $<$ 1.05 are in close pairs, 
and thus were likely triggered by galaxy interactions.
\citet{C11} argue that AGN are as likely as quiescent galaxies to show signs of a recent merger at $z$ $<$ 1, and thus mergers cannot 
be connected to AGN activity at this redshift range.  \citet{G09} investigate the morphology of AGN hosts and find that AGN hosted 
by disk galaxies contribute $\sim$25$\%$ of the AGN luminosity density at $z$ $\approx$ 0.8.  They conclude, therefore, that a large 
fraction of $z$ $\sim$ 1 AGN are triggered by mechanisms unrelated to major mergers.  \citet{C10} found that when AGN host galaxy 
colors are corrected for dust extinction, there is a bi-modality of AGN host galaxy colors at $z$ $\sim$ 1, suggesting that there 
are two modes of AGN activity at this redshift range.  When modeling the AGN population by considering mergers of massive dark matter 
halos, \citet{Shen09} find that secular AGN activity is necessary at $z$ $<$ 0.5 to account for the observed AGN population.  \citet{K10} 
recently found that $\sim$20$\%$ of {\em Swift}/BAT AGN at $z$ $<$ 0.05 are hosted by galaxies with disturbed morphologies indicative of 
a recent major merger and an additional 6$\%$ of {\em Swift}/BAT AGN are in close pairs suggesting these AGN were triggered by galaxy interactions.
Thus, it appears that both major mergers and 
secular processes must contribute to the AGN activity observed at $z$ $\lesssim$ 2.  If secular triggers dominate the AGN 
population at $z$ $\lesssim$ 2, current understanding of the stochastic fueling of secular mechanisms is insufficient to explain the 
high luminosity of quasars observed at $z$ $\lesssim$ 2 \citep{HH06}.  However, \citet{Bo11} recently showed that, at least at high 
redshift, it is possible for cold-flow accretion to trigger moderate luminosity AGN with occasional bright episodes.  Thus, 
phenomenological models of the AGN population may provide an important tool for 
understanding the importance of merger and secularly triggered AGN at various redshift and luminosity ranges.

Here, a model of the AGN population is calculated using an empirically motivated space density of AGN triggered at each redshift 
and a theoretical AGN light curve.  We consider whether major mergers or secular mechanisms alone can account for the observed 
AGN hard X-ray luminosity function (HXLF), AGN number counts, black hole space density, and the X-ray background (XRB) spectrum.  
In Section \ref{sect:model} the details of the AGN model, including the triggering rate, light curve, and black hole mass functions 
utilized, are described.  Section \ref{sect:calc} explains the calculations completed to compare the model against observations.  We 
then consider the model results if AGN are a single population (Section \ref{sect:one}) or two populations (Section \ref{sect:two}).  
The results are discussed and summarized in Sections \ref{sect:disc} and \ref{sect:sum}.  A $\Lambda$CDM cosmology is assumed with H$_0$ = 70 
km s$^{-1}$ Mpc$^{-1}$ and $\Omega_{\Lambda}$ = 1.0 - $\Omega_{M}$ = 0.7.

\section{The AGN Population Model}
\label{sect:model}

In order to model the evolution of the AGN HXLF, three ingredients are necessary.  
To determine the space density of triggered AGN, an empirically based major merger rate is used.  Once 
triggered, the AGN Eddington ratio, $\lambda$ = $L_{bol}/L_{Edd}$, where $L_{bol}$ is the bolometric luminosity 
and $L_{Edd}$ is the Eddington luminosity, is evolved using a theoretically motivated light curve. The active 
black hole mass function (ABHMF) and the \citet{M04} bolometric correction are then used to convert from Eddington 
ratio to 2--10 keV luminosity, $L_X$.  These three ingredients are described in detail below.

\subsection{Triggering Rate}
\label{sub:trigger}

In order to determine the space density of AGN triggered at each redshift, the space density of gas rich 
massive galaxies at redshift $z$ must be calculated.  First, the minimum stellar mass of a massive galaxy, $M_*^{min}$, 
at $z$ is derived by parametrizing the median mass of ultra-luminous infrared galaxies (ULIRGS) as a function of redshift \citep{T10}, which gives
\begin{equation}
M_*^{min}(z) = 5 \times 10^{11} (1.0+z)^{-1.5} M_{\odot}.
\label{eq:Mmin}
\end{equation}
The space density of massive galaxies at $z$, $N_{gal}(M_*>M_*^{min}(z))$, in Mpc$^{-3}$, can then be calculated by 
integrating the stellar mass function from 
$M_*^{min}(z)$ to $M_*^{max}$ = 10$^{12.5}$ M$_{\odot}$, such that
\begin{equation}
N_{gal}(M_*>M_*^{min}(z)) = \int_{M_*^{min}(z)}^{M_*^{max}} \frac{d\Phi_{gal}(M_*,z)}{d\log M_*}d \log M_*,
\label{eq:Ngal}
\end{equation}
where $d\Phi_{gal}/d\log M_*$ is the stellar mass function (SMF) of \citet{PG08} for $z$ $\lesssim$ 4.  The dependence 
of the results on the SMF is considered in Sections \ref{sub:merg} and \ref{sub:sec}. 
To determine the space density of gas rich massive galaxies, the fraction of gas rich galaxies at $z$, $f_g(z)$, 
must be determined.  Considering observations of the GOODS fields \citep{Dah07}, \citet{T10} find that $f_g(z)$ 
can be parametrized as
\begin{equation}
f_g(z) = \left\{ \begin{array}{l l}
0.11(1+z)^{2.0}  & z\leq 2 \\ 
1 & z>2
\end{array}\right..
\label{eq:fg}
\end{equation}
Thus the space density, in Mpc$^{-3}$, of potential AGN host galaxies at redshift $z$ is $f_g(z)N_{gal}(M_*>M_*^{min}(z))$.

The space density of AGN triggered by a merger at redshift $z$ is then calculated by multiplying the space density of 
potential AGN host galaxies at $z$ by the fraction of massive galaxies which will undergo a merger at $z$.  Following 
\citet{H10}, who derive the major merger rate per galaxy per Gyr, $d^2\Psi$/$dt$ $dN$, from simulations and observational 
constraints, we parametrize $d^2\Psi$/$dt$ $dN$ as
\begin{equation}
\frac{d^2 \Psi}{dt\,dN} = A(M_*^{min})(1.0+z)^{\beta(M_*^{min})},
\label{eq:dmerg}
\end{equation}
where 
\begin{equation}
A(M_*) = 0.02\left[1+\left(\frac{M_*}{2\times 10^{10} M_{\odot}}\right)^{1/2}\right] \textnormal{Gyr}^{-1}
\label{eq:A}
\end{equation}
and 
\begin{equation}
\beta(M_*) = 1.65 - 0.15\log\left(\frac{M_*}{2\times 10^{10} M_{\odot}}\right).  
\label{eq:beta}
\end{equation}
Here the major merger is said to occur upon the coalescence of the two similarly massive galaxies \citep{H10b}.  
By assuming that every major merger 
leads to an AGN event, with a negligible time delay between the merger and the triggering of the AGN activity, the 
space density of AGN triggered by major mergers, $dN_{merg}$, at redshift $z$ is
\begin{equation}
dN_{merg} (z) = \frac{d^2 \Psi}{dN dt} \,N_{gal}(M_* > M_*^{min}(z)) \,f_{g}(z)\, dt \, \textnormal{Mpc}^{-3}.
\label{eq:mergrate}
\end{equation}
%

The rate at which AGN are triggered through secular processes, such as galaxy interactions, cold gas accretion, 
and internal disk instabilities, is calculated in a similar manner.  For AGN triggered by secular processes, 
$M_*^{min}(z)$ = 5 $\times$ 10$^9$ M$_{\odot}$, in agreement with the findings of \citet{S10}.
The fractional rate of massive gas rich galaxies which are triggered through secular processes every Gyr, $f_{sec}$, is 
assumed to be constant with redshift.  \citet{Y09} found that the fraction of galaxies with $M_*$ $>$ 10$^{11}$ M$_{\odot}$ 
hosting AGN, which are triggered through both secular processes and mergers, is $\sim$0.3.  As the fraction of AGN triggered 
by secular processes is poorly constrained observationally, and we are  
considering galaxies with $M_*$ $>$ 5 $\times$ 10$^9$ M$_{\odot}$, the constraint $f_{sec}$ $\ll$ 0.3 Gyr$^{-1}$ is used.  
The specific value of $f_{sec}$ is set by calculating the predicted AGN HXLF and minimizing a $\chi^2$ test which compares against 
the observed HXLF.  The space density of AGN triggered by secular 
mechanisms, $dN_{sec}$, at redshift $z$ is then
\begin{equation}
dN_{sec} (z) = f_{sec} \,N_{gal}(M_* > M_*^{min}(z)) \,f_{g}(z)\, dt \, \textnormal{Mpc}^{-3}.
\label{eq:secrate}
\end{equation}
%

\subsection{AGN Light Curve}
\label{sub:lc}

Once the AGN has been triggered, its Eddington ratio is used to 
parametrize the accretion as a function of time since the 
AGN was triggered.  While there is evidence that AGN are an intermittent phenomenon, 
\citet{T10} show that quasars can grow most of their black hole mass in a single, merger triggered event.  Therefore, a 
single-peaked light curve is assumed.  

Based on hydrodynamical simulations, \citet{HH09} suggest 
\begin{equation}
\lambda(t) = \left[1+\left(\vert t\vert/t_Q\right)^{1/2}\right]^{-2/\beta},
\label{eq:lc}
\end{equation}
where $t = t_{on} - t_Q$, where $t_{on}$ is the time since the AGN was triggered, $t_Q$ = $t_0 \eta^{\beta}/(2\beta \ln 10)$, 
and $t_0$, $\eta$, and $\beta$ are fitting parameters which describe the quasar lifetime, maximum Eddington ratio, and light 
curve slope, respectively\footnote{For completeness, a wide variety of light curves were tested, including a constant $\lambda$, 
a monotonic linearly increasing or decreasing $\lambda$, a 
monotonic exponentially increasing or decreasing $\lambda$, a linearly increasing followed by a linearly decreasing $\lambda$, 
and an exponentially increasing followed by an exponentially decreasing $\lambda$.  However, these light curve models were 
unable to match the shape of the AGN HXLF.}.  

\subsection{Active Black Hole Mass Function and Its Evolution}
\label{sub:abhmf}

By combining the triggering rate and light curve, the space density of AGN with Eddington ratio $\lambda$ can be calculated at any 
redshift $z$ $\lesssim$ 4.  However, to compute model predictions that can be compared to observational constraints, the black hole mass 
must be used to convert Eddington ratios into bolometric luminosities.  Once the bolometric luminosity is computed, the \citet{M04} 
bolometric correction is used to determine $L_X$.  The distribution of black hole masses is determined by the fractional active black hole mass 
function (ABHMF) at $z$, which describes the fraction of active black holes at redshift $z$ with black hole mass $M_{\bullet}$.

Two ABHMFs are considered.  The first ABHMF is a Gaussian fit to the combined type 1 and type 2 AGN ABHMF at $z$ $\sim$ 0.15 observed 
by \citet{N09}.  The second ABHMF considered is a Schechter function with the same slope and critical mass, $M_{crit}$, as the black hole 
mass function described by \citet{MH08}.  Both ABHMFs are considered over the range $\log (M_{\bullet}^{min}/M_{\odot})$ = 5.95 and 
$\log (M_{\bullet}^{max}/M_{\odot})$ = 10.55.  A black hole with $M_{\bullet}$ $<$ $M_{\bullet}^{min}$ would 
need to accrete at Eddington ratio $\lambda$ $\gtrsim$ 0.1 to achieve $\log L_X$ $\gtrsim$ 41.5.  According 
to the light curve used here, the vast majority of the AGN lifetime is spent at $\lambda$ $<$ 0.1.
Therefore, black holes with $M_{\bullet}$ $<$ $M_{\bullet}^{min}$ 
are expected to only make a very small contribution to the observed AGN population \citep{M04, MH08}.  Black holes with $M_{\bullet}$ $>$ 
$M_{\bullet}^{max}$, correspond to $\ll$ 0.1$\%$ of all active black holes according to both the \citet{MH08} 
and \citet{N09} ABHMFs, in agreement with the maximum black hole mass derived by \citet{NT09}. 

Both the \citet{MH08} and \citet{N09} ABHMFs are derived using scaling relationships to determine black hole masses.  
These scaling relationships have an intrinsic scatter of $\sim$0.3 dex \citep{PB06,MH08, N09}.  This leads to an 
uncertainty in the calculated $L_X$ of a factor $\sim$ 2 $\times$ $\lambda$.  For the majority of sources, $\lambda$ $<$ 0.1, 
thus the uncertainty introduced by the ABHMF is a smaller part of the error budget than the uncertainty in the major merger rate, 
which is a factor $\sim$2.  

For both ABHMFs two redshift evolutions are investigated.  The first ABHMF evolution uses the continuity equation \citep{SB92, MH08}
\begin{equation}
\frac{\partial n_M(M_{\bullet},t)}{\partial t}+\frac{\partial[n_M(M_{\bullet},t)\langle\dot{M}(M_{\bullet},t)\rangle]}{\partial M}=0,
\label{eq:coneq}
\end{equation}
where $n_M$ is the ABHMF and $\langle\dot{M}(M_{\bullet},t)\rangle$ is the average accretion rate of black holes with mass $M_{\bullet}$ 
at time $t$.  By integrating this 
conservation equation forward and backward in time, using the observed ABHMF as the boundary condition, the ABHMF can 
be evolved to any redshift $z$.  This evolution assumes that black hole growth occurs through accretion and that binary mergers are 
not the primary mechanism of black hole growth \citep[e.g.,][]{V03}.  The second ABHMF evolution considered is based on the 
observations of \citet{L09} who found that the 
maximum black hole mass of the quasar population increases with redshift.  Thus, $M_{\bullet}^{crit}(z)$ = 
$M_{\bullet}^{crit}(0)(1.0+z)^{1.64}$.  In order to compare the two considered evolutions, a power law is fit to 
$M_{\bullet}^{crit}(z)$ of the \citet{MH08} ABHMF evolved with the continuity 
equation and it is found that $M_{\bullet}^{crit}(z)$ $\approx$ $M_{\bullet}^{crit}(0)(1.0+z)^{0.5}$.    
For both ABHMFs, the ABHMF is re-normalized at each redshift so that integrating over all black hole 
masses gives 1.

With the observationally determined AGN triggering rate, theoretical Eddington ratio evolution, and the ABHMF in place, the 
AGN population at $z$ $<$ 4 can be fully 
modeled\footnote{AGN triggered at $z$ $>$ 4 make only a minor contribution to the $z$ $\lesssim$ 2 AGN population.  
Therefore, if the AGN triggered at $z$ $>$ 4 are included, the results of this study do not change.}.  
To determine the light curve fitting parameters, $t_0$, $\beta$, and $\eta$, 
the AGN population model must be compared to observational constraints.


\section{Calculations and Observational Constraints}
\label{sect:calc}

In order to constrain the light curve parameters, the AGN population model is compared against the HXLF at five different redshifts, the black hole mass density as 
a function of redshift, the {\em Swift/BAT} 15--55 keV AGN counts, the 2--10 keV AGN counts, and the XRB 
spectrum.  The methods used to calculate these quantities are described below.

\subsection{AGN HXLF}
\label{sub:HXLF}

The HXLF, $d\Phi_X (L_X,z)/d\log L_X$, is computed at five different redshifts, $z$ = 2.3, 1.2, 0.6, 0.3, and 0.1.
The HXLFs observed by \citet{U03}, \citet{LF05}, \citet{S08}, \citet{Aird10}, and \citet{U11} are each presented in 
different redshift bins.  Therefore, we compute the predicted HXLF at the central redshift of each of the \citet{U03} 
redshift bins.

The bolometric AGN luminosity function at redshift $z$, $d\Phi (L_{bol},z)/d\log L$, is computed by integrating over the space density of AGN triggered at redshift $z_t$ and, at redshift $z$, have black hole mass $M_{\bullet}$ and Eddington ratio $\lambda$ such that the bolometric luminosity is $L_{bol}$.  Thus,
\begin{equation}
\frac{d\Phi(L_{bol}, z)}{d\log L} = \int_{M_{\bullet}^{min}}^{M_{\bullet}^{max}} n_{M}(M_{\bullet},z) d\log M_{\bullet} \int^4_z \Phi_{\lambda}(M_{\bullet},z) dz_t,
\label{eq:blf}
\end{equation}
where $\log (M_{\bullet}^{min}/M_{\odot})$ = 5.95, $\log (M_{\bullet}^{max}/M_{\odot})$ = 10.55, $z_t$ is the triggering redshift, and 
\begin{equation}
\Phi_{\lambda}(M_{\bullet},z) = \left\{ \begin{array}{l l}
\frac{dN(z_t)}{d\log L} & \lambda(z,z_t) L_{Edd}(M_{\bullet}) = L_{bol} \\
0 & \textnormal{otherwise}
\end{array}\right.,
\label{eq:philambda}
\end{equation}
where $dN$ is either $dN_{merg}$ or $dN_{sec}$.  The \citet{M04} 
luminosity dependent bolometric correction is then used to convert $L_{bol}$ to $L_X$.  

As the observed HXLF only includes Compton thin AGN, the Compton thick AGN are removed from the calculated HXLF by assuming that one-third of 
all obscured AGN are Compton thick in agreement with the fraction of Compton thick AGN necessary for the \citet{U03} observed HXLF to 
be in agreement with the peak of the XRB at $\sim$ 30 keV \citep{DB09, DB10, B11}.  The fraction of Compton thin obscured sources, $f_2$, 
is assumed to be a function of luminosity and redshift such that $f_2 \propto (1+z)^a(\log L_X)^{-b}$, where $a$ = 0.4 \citep{B06,TU06} 
and $b$ = 4.7.  The normalization factor is determined by assuming the type 2 to type 1 AGN ratio is 4:1 at $z$ = 0 and $\log L_X$ = 41.5.  
Thus, the space density of Compton thick AGN also depends on $L_X$ and, in Mpc$^{-3}$ dex$^{-1}$, is $(f_2/2)$ $d\Phi_X/d\log L_X$.

According to both the \citet{M04} and \citet{V09} bolometric corrections, to achieve $L_X$ = 10$^{46.5}$ erg s$^{-1}$ it 
is necessary to have a black hole with mass $\approx$10$^{10.7}$ -- 10$^{10.9}$ M$_{\odot}$ accreting at its Eddington rate.   
Both the Gaussian fit to the \citet{N09} and the \citet{MH08} ABHMF, predict the fraction of active black holes with mass 
$\sim$10$^{11}$ M$_{\odot}$ is $\sim$ 0, in agreement with the black hole mass upper limit derived by \citet{NT09}.  This 
strongly suggests that AGN with $L_X$ $\gtrsim$ 10$^{46}$ erg s$^{-1}$ are accreting at super-Eddington rates.  The light 
curve model used allows for 0 $\lesssim$ $\lambda$ $\lesssim$ 1.  Therefore, super-Eddington accretion cannot be taken into 
account.  Thus, the models presented here will necessarily under-predict the $L_X$ $>$ 10$^{46}$ erg s$^{-1}$ observed HXLF 
data points.  For completeness, the HXLF data points at $L_X$ $>$ 10$^{46}$ erg s$^{-1}$ are still taken into account when 
performing the $\chi^2$ fitting\footnote{Exclusion of the three data points at $L_X$ $>$ 10$^{46}$ erg s$^{-1}$ changes 
the reduced $\chi^2$ values by $\lesssim$ 0.1 and does not affect the determination of the light curve parameters.}.

\subsection{Black Hole Mass Density}
\label{sub:massden}

To calculate the black hole mass density, the \citet{S82} argument is used.  Thus, the black hole mass density at redshift $z$, 
$\rho_{\bullet}(z)$, is calculated as
\begin{equation}
\rho_{\bullet} (z) = \int_z^{\infty}\frac{dt}{dz}dz\int_0^{\infty}\frac{1-\epsilon}{\epsilon c^2}L_{bol}\frac{d\Phi(L,z)}{d\log L}d\log L
\label{eq:soltan}
\end{equation}
where $\epsilon$ = 0.1 is the radiative efficiency, $c$ is the speed of light, and $d\Phi(L,z)/d \log L$ is the AGN luminosity function 
\citep{YT02, M04, T11} \footnote{It is expected that at Eddington ratios $\lambda$ $\lesssim$ 10$^{-2}$, AGN accretion flows become 
radiatively inefficient and the radiative efficiency decreases with $\lambda$ \citep[e.g.,][]{CX07}.  If Equation 1 of \citet{MH08} 
is used to characterize $\epsilon$, the calculated 
local black hole mass density changes by $<$ 5$\%$ since the majority of black hole growth occurs during radiatively efficient accretion.}.  
The contribution of Compton thick AGN to the black hole mass density is included.  
The predicted black hole mass density is compared to the 
local black hole mass density observed by \citet{S09} and the $z$ $\sim$ 2 black hole mass density observed by \citet{T10b}.

%
%

\subsection{X-ray Background Spectrum}
\label{sub:xrb}

The XRB spectrum model closely follows that described by \citet{DB09}.  Instead of inputting the observed luminosity function, 
the HXLF calculated as described in Section \ref{sub:HXLF} is used.  Also, a torus reflection component \citep{G07} is included 
which is computed using "reflion" \citep{RF05}.  The Type 2 fraction, $f_2$, is determined as in Section \ref{sub:HXLF}.
The Compton thick fraction, $f_{CT}$, is defined as the ratio of the number of Compton thick AGN to the number of Compton 
thin type 2 AGN, and is set to $f_{CT}$ = 0.5, in agreement with the 
$f_{CT}$ necessary for the \citet{U03} HXLF to be in agreement with the XRB spectrum and the local Compton thick AGN space 
density \citep{DB09, DB10, B11}.  The unabsorbed type 1 sources are evenly distributed over column densities 
$\log N_{H}$ = 20.0, 20.5, 21.0, and 21.5.  The Compton thin type 2 sources are distributed equally over 
$\log N_{H}$ = 22.0, 22.5, 23.0, and 23.5.  The contribution of Compton thick AGN to the XRB is included.  To do so, it is 
assumed that Compton thick sources evolve similarly to 
less obscured AGN and are evenly distributed over $\log N_{H}$ = 24.0, 24.5, and 25.0.  

The AGN number counts in the 2--10 keV and 15--55 keV bands are also calculated.  This is done by using the same AGN spectra, 
$f_2$, and $f_{CT}$ as in the XRB model described above, including the contribution of Compton thick AGN, and by using the HXLF 
calculated as in Section \ref{sub:HXLF}.

\subsection{Summary of Free Parameters}
\label{sub:freeparam}

The light curve fitting parameters--- $t_0$, which is related to the AGN lifetime, $\eta$, 
which is related to the peak Eddington ratio, and $\beta$, which determines the slope of the light curve--- are determined by 
comparing the resulting models against the observed HXLF, evolving black hole space density, XRB spectrum, and AGN number counts.  
The fractional rate of massive galaxies which are triggered by secular processes each Gyr, $f_{sec}$, is also determined by comparing against 
observational constraints.  

The focus of this study is not to fit the AGN light curve parameters, but to investigate the contribution to the AGN population of AGN 
triggered by major mergers and AGN triggered by secular processes.  Therefore, a minimization algorithm is not used.  Instead, 
the best fit parameters for the AGN light curve are determined by considering models with $t_0$ = 1.0 $\times$ 10$^7$, 
5.0 $\times$ 10$^7$, 1.0 $\times$ 10$^8$, 2.5 $\times$ 10$^8$, 5.0 $\times$ 10$^8$, 7.5 $\times$ 10$^8$, and 1.0 $\times$ 
10$^9$ yrs.  \citet{HH06} found that their model results were not very dependent on $\eta$, so the values $\eta$ = 0.2, 0.4, 1.0, 2.5, 
and 3.0 are considered.  Steps of 0.05 in the range from 0.05 to 
1.0 are used to determine $\beta$.  For $f_{sec}$ the values 0.0, 0.005, 0.01, 0.02, and 0.05 Gyr$^{-1}$ are 
investigated.  The best fit parameters for each model are determined by minimizing a $\chi^2$ test which takes into account the 127 HXLF 
data points for $z$ = 2.3, 1.2, 0.6, 0.3, and 0.1 presented by \citet{U03}, \citet{LF05}, \citet{S08}, \citet{Aird10}, and \citet{U11}.  
By comparing against the observed AGN HXLF, black hole mass density, XRB spectrum, and AGN number counts, the model is fully constrained, allowing for some conclusions to be reached about the AGN light curve parameters.  The model sensitivity to the light curve parameters is discussed in Section \ref{sub:lcm}.

\section{One Population}
\label{sect:one}

\subsection{Major Merger Triggered Quasars}
\label{sub:merg}

\begin{figure*}
\begin{center}
\includegraphics[angle=0,width=0.95\textwidth]{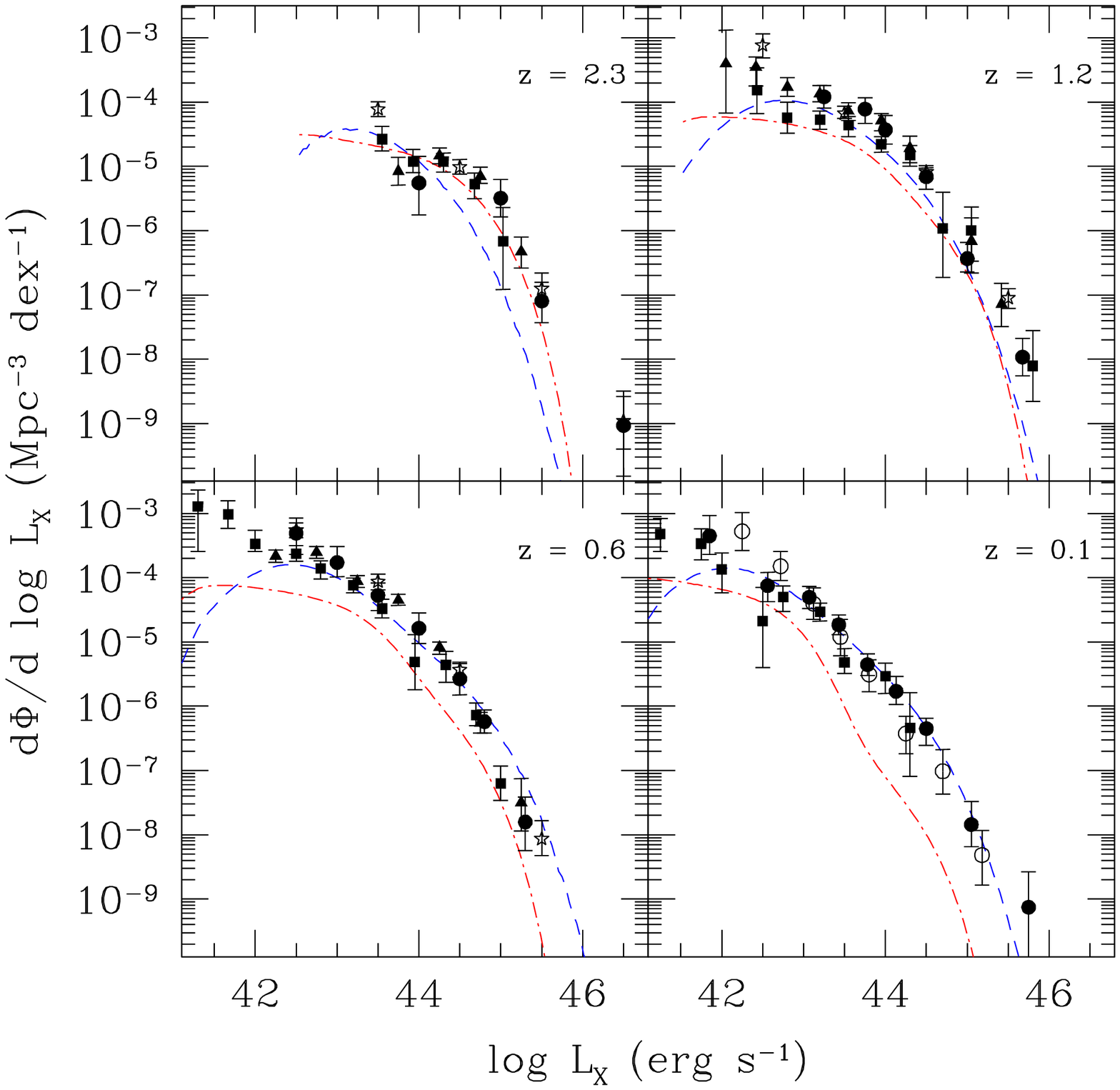}
\end{center}
\caption{Best fit one population models compared against the observed HXLF.  The dot-dashed red lines show the HXLF from the best fit major merger trigger only model and the dashed blue lines show the HXLF from the best fit secular processes trigger only model.  The data points show measurements of the HXLF by Ueda et al. (2003; filled circles), La Franca et al. (2005; stars), Silverman et al. (2008; triangles), Aird et al. (2010; squares), and Ueda et al. (2011; open circles).}
\label{fig:one}
\end{figure*}

First, it is assumed that all AGN are triggered by major mergers, thus, $f_{sec}$ = 0.0.  The best fit to the HXLF is found using the \citet{MH08} 
ABHMF with the \citet{L09} evolution, $t_0$ = 2.5 $\times$ 10$^{8}$ yrs, $\beta$ = 0.7, and $\eta$ = 2.5, similar to that found by \citet{Cao10}.  
When compared to the 127 data points from the \citet{U03}, \citet{LF05}, \citet{S08}, \citet{Aird10}, and \citet{U11} observed HXLFs in the 5 
redshift ranges considered, this model has a reduced $\chi^2$, $\chi^2_{red}$ = 2.4.  If $t_0$ is changed by 0.5 $\times$ 10$^8$ yrs, 
$\beta$ is changed by 0.05, or $\eta$ is changed by 0.5, the resulting $\chi^2_{red}$ will increase by $\sim$ 0.1 -- 0.2.  The light curve 
fitting parameters are summarized in Table \ref{modelparam}.  As shown by the dot-dashed red lines in Figure \ref{fig:one}, 
this model provides a relatively good fit to the observed HXLF at $z$ $\gtrsim$ 1; however, at $z$ $<$ 1, this model has $\chi^2_{red}$ 
= 2.6 for 85 data points.  The space density of major mergers at $z$ $<$ 1 is too low to explain the space density of observed AGN in this redshift range.  
This model is in agreement with the observed local black hole mass density \citep{S09} and the $z$ $\sim$ 2 observed black hole mass 
density \citep{T10b}.  However, this model significantly under-predicts the XRB spectrum as well as the 2--10 keV and 15--55 keV AGN number counts.  
Thus, despite major merger triggered AGN being able to account for the integrated black hole growth, merger triggered AGN cannot account for the 
space density of the entire AGN population.  
Specifically, major mergers cannot account for the $z$ $\lesssim$ 1 AGN population.

The finding that major mergers are not capable of accounting for the $z$ $\lesssim$ 1 AGN population is robust against several assumptions.  
To test the dependence of the merger triggered AGN only model on the SMF, the $z$ = 1.3--3 SMF presented 
by \citet{March09} and the $z$ = 0.1 SMF presented by \citet{C01} are used to define an evolving SMF.  When using the combined \citet{C01} and 
\citet{March09} SMF, the merger triggered AGN only model is not able to account for the space density of AGN at $z$ $\lesssim$ 1, with minimum $\chi^2_{red}$ = 3.6.  Similarly, if the 
minimum mass of a potential AGN host galaxy ($M_*^{min}$) is reduced, the merger only model cannot supply a decent fit to the observational constraints.


\subsection{Secularly Triggered AGN}
\label{sub:sec}

Next, we consider the scenario where all AGN are triggered by secular processes.  For this model, the best fit to the HXLF is found 
when using the \citet{N09} ABHMF with the continuity equation evolution, $t_0$ = 2.5 $\times$ 10$^{8}$ yrs, $\beta$ = 0.8, $\eta$ 
= 0.4, and $f_{sec}$ = 0.02 Gyr$^{-1}$.  This model has $\chi^2_{red}$ = 2.1 for 127 data points.  If $t_0$ is changed by $\sim$ 
0.5 $\times$ 10$^8$ yrs,$\beta$ is changed by 0.05, or $f_{sec}$ is changed by 0.01 Gyr$^{-1}$, the resulting $\chi^2_{red}$ will 
change by $\sim$ 0.2.  If $\eta$ is changed 
by $\sim$ 0.5 the resulting $\chi^2_{red}$ will increase by $\sim$ 0.1.  As shown by the blue 
dashed lines in Figure \ref{fig:one}, this model under-predicts the space density of low $L_X$ AGN at $z$ $\lesssim$ 1 and under-predicts 
high $L_X$ sources at $z$ $\gtrsim$ 1.    
Furthermore, this model significantly under-predicts the local black hole mass density, the XRB, and the 2--10 keV number counts.  This model 
slightly over-predicts the {\em Swift}/BAT 15--55 keV number count and is inconsistent with the findings of \citet{K10} that 
$\sim$20$\%$ of {\em Swift}/BAT AGN host galaxies have disturbed morphology indicative of a recent major merger.  
Thus, secular processes alone are not sufficient to account for the entire AGN population.  Specifically, secular processes are not 
able to account for the AGN population at $z$ $\gtrsim$ 0.5.  When the minimum mass of a potential AGN host galaxy is reduced or the 
evolving SMF defined by combining the \citet{C01} and \citet{March09} SMFs is used, the secular evolution only model still cannot account 
for the $z$ $\gtrsim$ 0.5 AGN population, with minimum $\chi^2_{red}$ = 2.0.  It is found that AGN triggered by 
secular mechanisms alone cannot explain the observed AGN population.



Neither merger triggered AGN nor secularly triggered AGN alone are able to account for the entire AGN 
population.  Reducing the minimum mass of potential AGN host galaxies and using the combined \citet{C01} 
and \citet{M09} SMF does not change this result.  The space density of major mergers at $z$ $<$ 1 is too 
low to account for the low redshift AGN population but can account for the local space density of black 
holes.  In contrast, AGN triggered by secular mechanisms are not able to account for the build up of the 
black hole mass density over cosmic time.  Thus, we consider the possibility that AGN are triggered by 
both major mergers and secular processes.

\section{Two Populations}
\label{sect:two}

As neither the major merger triggered AGN nor the secularly triggered AGN can alone account for the entire AGN population, 
we now consider a two population model in which AGN are assumed to be triggered by both secular processes and major 
mergers.  First, the major merger triggered AGN light curve is determined by finding the lowest $\chi^2$ fit to the $z$ = 2.3 HXLF 
using only the major merger triggered portion of the AGN population.  Thus, the maximum possible contribution from major mergers is 
assumed\footnote{If the maximum contribution from secularly triggered AGN is assumed, an appropriate fit to the HXLF can be found ($\chi^2_{red}$ = 1.7), but this model cannot account for the observed local black hole mass density.}.  The light curve of the secularly triggered AGN 
is constrained by the minimum total $\chi^2$ fit to all 127 HXLF data points.  Both the \citet{MH08} and \citet{N09} ABHMFs and 
the \citet{L09} and continuity equation (Eq. \ref{eq:coneq}) ABHMF evolutions are 
considered.  It is assumed that merger triggered and secularly triggered AGN may have different light curves; thus, the light curve 
parameters for AGN triggered by mergers and AGN triggered by secular processes are considered separately.   

We started by attempting to model both the AGN triggered by major mergers and AGN triggered by secular processes using the same 
ABHMF and evolution for both populations of AGN.  The best fit parameters for these models are summarized by the middle 
section of Table \ref{modelparam}.  Despite these models providing decent fits to the observed HXLF data points, the models 
were not in agreement with the other observational constraints, as summarized by the middle section of Table \ref{modelfit}.  
However, this exercise did demonstrate that the $z$ $\lesssim$ 0.5 AGN population is fairly well described using the \citet{N09} 
ABHMF with the continuity equation evolution (Eq. \ref{eq:coneq}) and the $z$ $\gtrsim$ 1 AGN population can be described 
by using the \citet{MH08} ABHMF with the \citet{L09} evolution.  Therefore, a mixed ABHMF and evolution model was investigated.

\begin{figure*}
\begin{center}
\includegraphics[angle=0,width=0.95\textwidth]{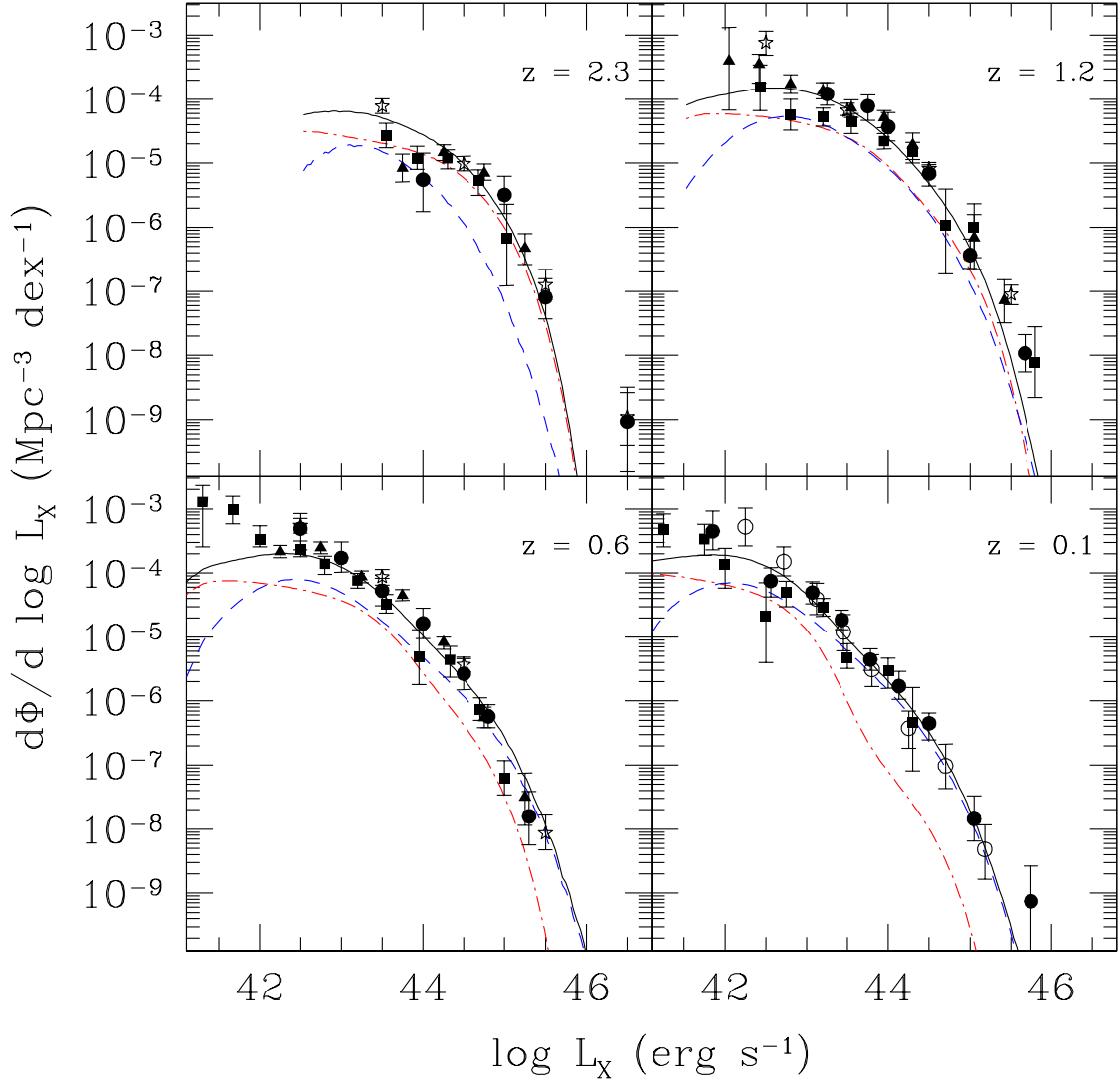}
\end{center}
\caption{The HXLF predicted by the mixed ABHMF model, which has $\chi^2_{red}$ = 1.4.  The solid black line is the total AGN HXLF.  The dot-dashed red lines show the contributions from AGN triggered by mergers and the dashed blue lines show the contribution from AGN triggered by secular mechanisms.  The data points are the same as in Figure \ref{fig:one}.}
\label{fig:mixedlf}
\end{figure*}

\begin{figure*}
\begin{center}
\includegraphics[angle=0,width=0.95\textwidth]{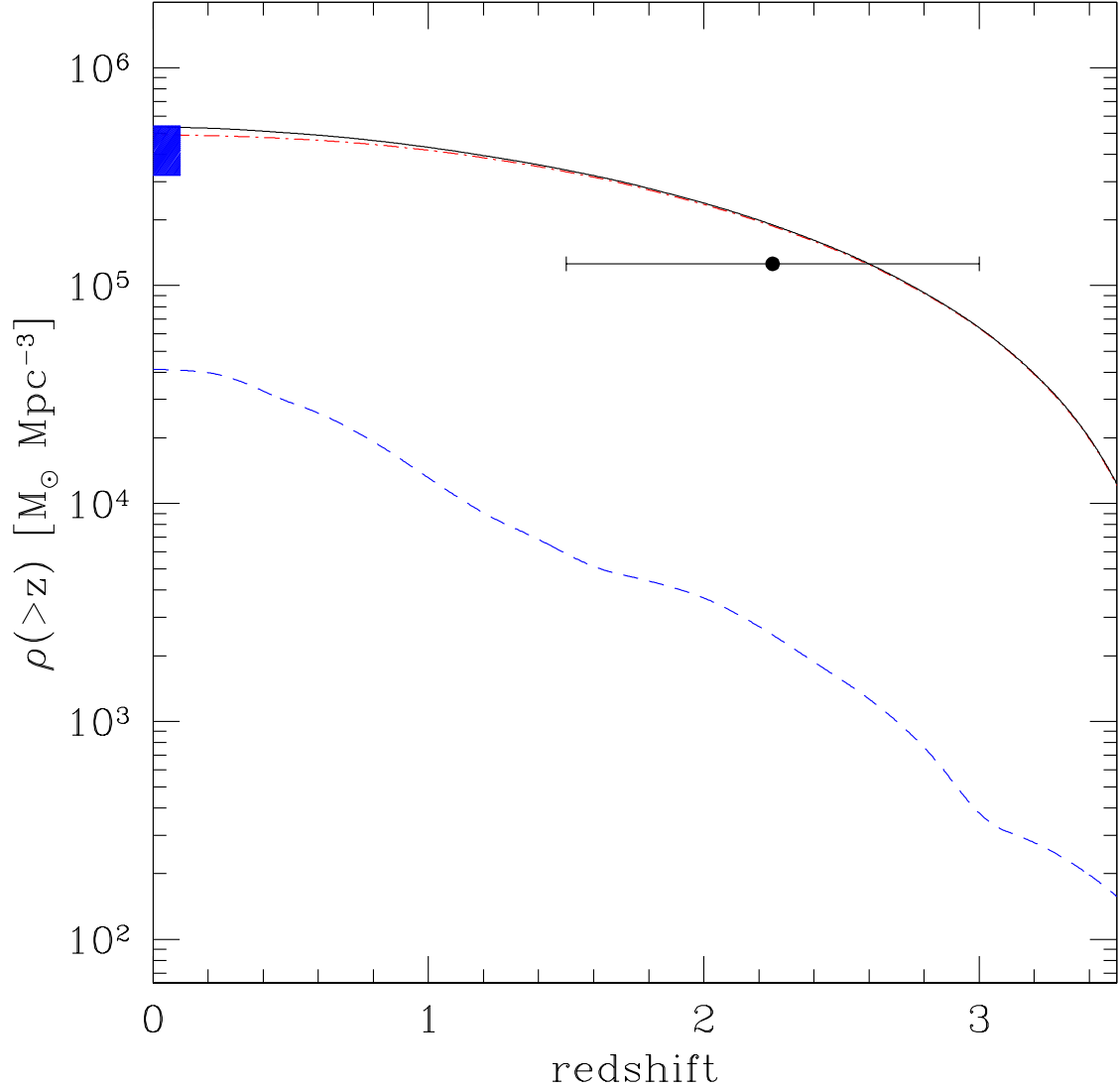}
\end{center}
\caption{The black hole mass density with respect to redshift, predicted by the mixed ABHMF model.  The line styles are the same as in Figure \ref{fig:mixedlf}.  The $z$ $\sim$ 0 shaded region is the local black hole mass density observed by \citet{S09} and the $z$ $\sim$ 2 data point is the black hole mass density observed by \citet{T10b}.}
\label{fig:mass}
\end{figure*}

\begin{figure*}
\begin{center}
\includegraphics[angle=0,width=0.95\textwidth]{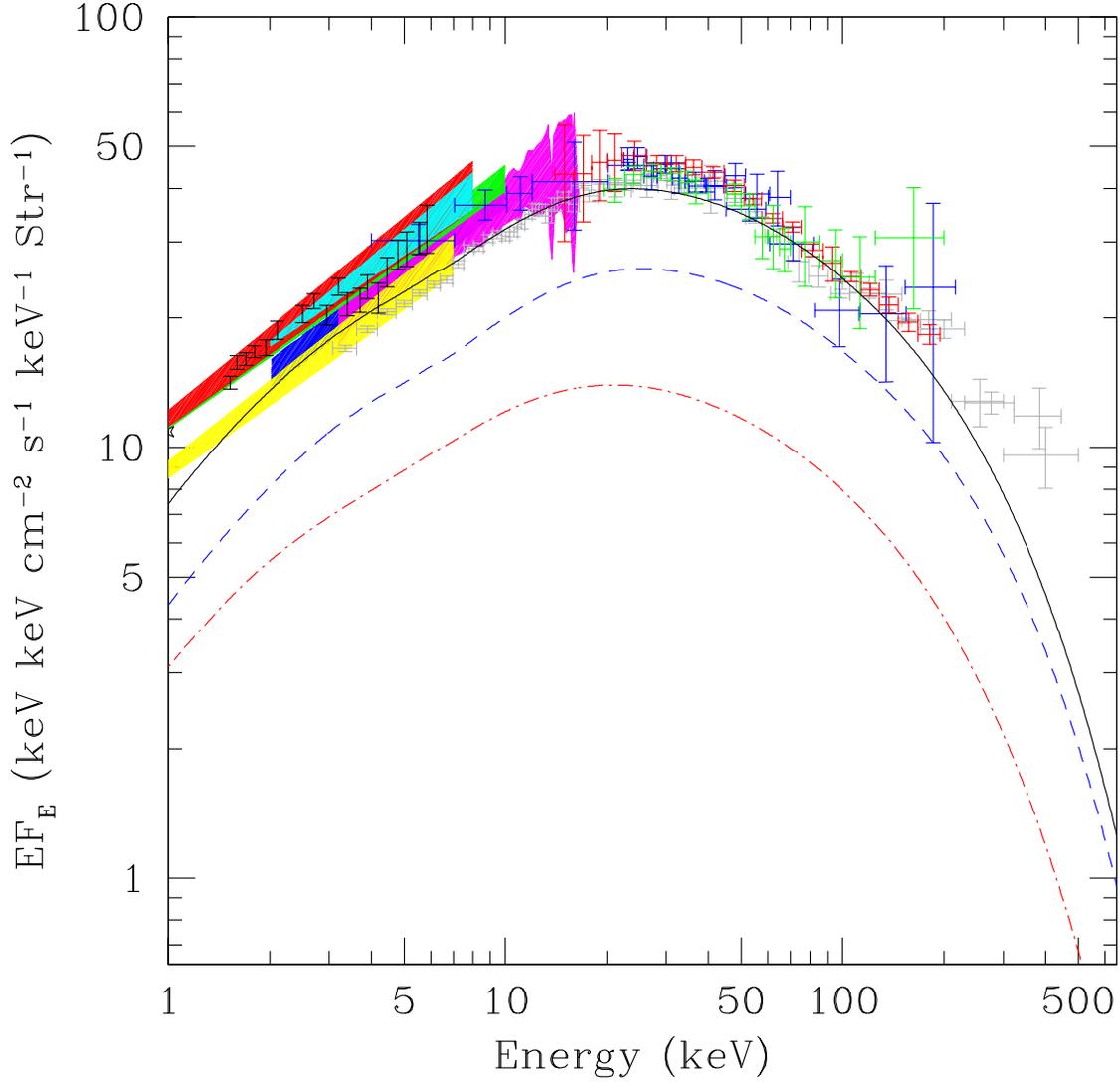}
\end{center}
\caption{The XRB spectrum predicted by the mixed ABHMF model.  The line styles are the same as in Figure \ref{fig:mixedlf}.  Colored regions and data points show measurements from various instruments; blue: {\em ASCA} GIS (Kushino et al. 2002); magenta: {\em Rossi X-ray Timing Explorer} ({\em RXTE}; Revnivtsev et al. 2003); green: XMM-{\em Newton} (Lumb et al. 2002); red: {\em BeppoSAX} (Vecchi et al. 1999); yellow: {\em ASCA SIS} (Gendreau et al. 1995); cyan: XMM-{\em Newton} (De Luca \& Molendi 2004); grey data: {\em HEAO}-1 (Gruber et al. 1999); blue data: {\em INTEGRAL} (Churazov et al. 2007); red data: {\em Swift}/BAT (Ajello et al. 2008); black data: {\em Swift}/XRT (Moretti \etal 2009); green data: {\em INTEGRAL} (T\"urler et al. 2010).}
\label{fig:xrb}
\end{figure*}

\begin{figure*}
\begin{center}
\includegraphics[angle=0,width=0.95\textwidth]{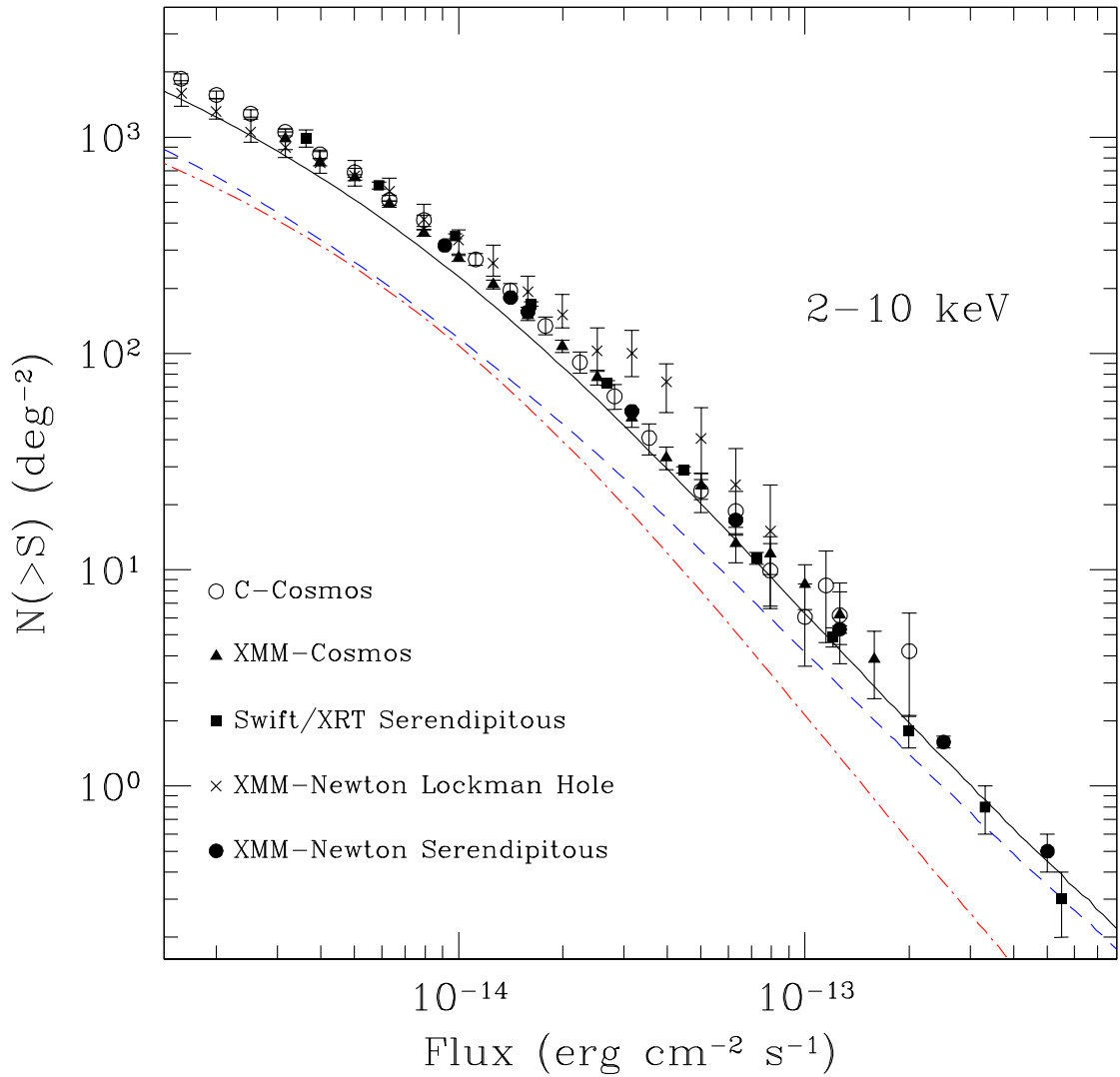}
\end{center}
\caption{The 2--10 keV AGN $N(>S)$ predicted by the mixed ABHMF model.  The line styles are the same as in Figure \ref{fig:mixedlf}.  The plotted data show observed the number counts from various surveys: C-COSMOS \citep{E09}, XMM-COSMOS \citep{Cap09}, {\em Swift}/XRT Serendipitous \citep{Puc11}, XMM-{\em Newton} Lockman Hole \citep{Br08}, and XMM-{\em Newton} Serendipitous \citep{M08}.}
\label{fig:2to8}
\end{figure*}

\begin{figure*}
\begin{center}
\includegraphics[angle=0,width=0.95\textwidth]{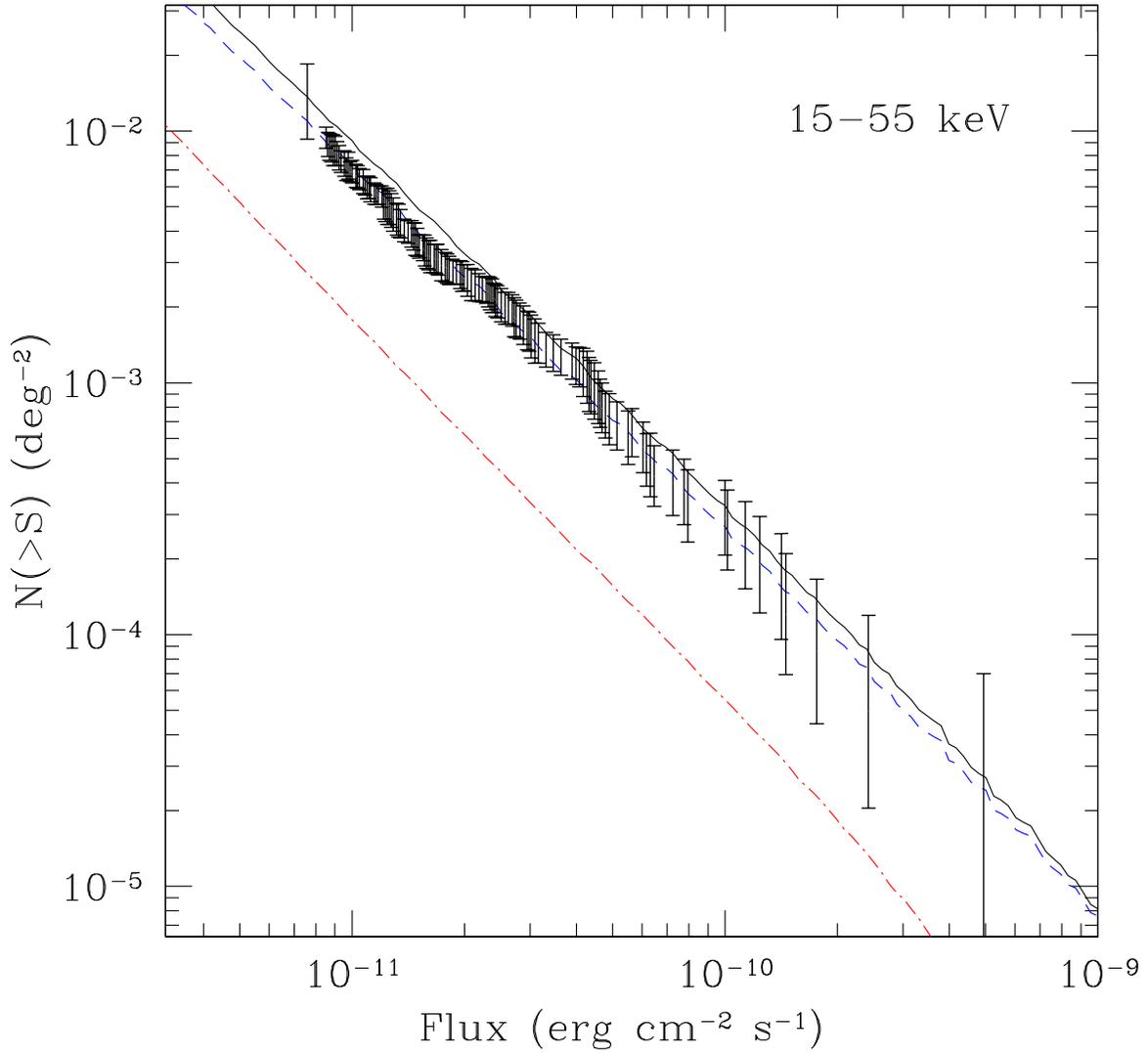}
\end{center}
\caption{The 15--55 keV AGN $N(>S)$ predicted by the mixed ABHMF model.  The line styles are the same as in Figure \ref{fig:mixedlf}.  The data points show the 15-55 keV AGN $\log N$--$\log S$ relation observed by {\em Swift}/BAT \citep{A09}.}
\label{fig:bat}
\end{figure*}

Since the \citet{N09} ABHMF with continuity equation evolution works well at low redshift, where secular evolution is expected to dominate, and the 
\citet{MH08} ABHMF with \citet{L09} evolution works well at high redshift, where major merger triggered AGN are expected to dominate, a hybrid 
model is calculated.  Using the \citet{MH08} ABHMF with the \citet{L09} evolution, the best fit parameters for the merger triggered AGN in this model are 
$t_0$ = 2.5 $\times$ 10$^{8}$ yrs, $\beta$ = 0.7, and $\eta$ = 2.5.  Using the \citet{N09} ABHMF with the continuity equation evolution (Eq. \ref{eq:coneq}), the AGN triggered by secular mechanisms are 
best described by the parameters $t_0$ = 2.5 $\times$ 10$^{8}$ yrs, $\beta$ = 0.8, $\eta$ = 0.4, and $f_{sec}$ = 
0.01 Gyr$^{-1}$.  This model does moderately well at describing the AGN HXLF, with $\chi^2_{red}$ = 1.4, as shown in Figure 
\ref{fig:mixedlf}.  Also, this model is in agreement with the local black hole mass density observed by 
\citet{S09} and the $z$ $\sim$ 2 black hole mass density observed by \citet{T10b}, which, as shown in Figure \ref{fig:mass}, 
is dominated by black hole growth triggered by major mergers.  
Figure \ref{fig:xrb} shows that this model is also in agreement with the observed XRB spectrum.  The mixed ABHMF 
model is in decent agreement with the 2--10 keV AGN number counts as shown in Figure 
\ref{fig:2to8}.  
Figure \ref{fig:bat} shows that the mixed ABHMF model is in good
agreement with the {\em Swift}/BAT 15--55 keV AGN number count observed by \citet{A09}.  Furthermore, at the {\em Swift}/BAT 
survey flux limit of 7.6 $\times$ 10$^{-12}$ erg cm$^{-2}$ s$^{-1}$, we find that merger triggered AGN account for $\sim$20$\%$ of 15--55 
keV number count, in agreement with the findings of \citet{K10}.  Thus, this mixed ABHMF model provides the best overall fit to the 
observational constraints.  

The largest discrepancy between the mixed ABHMF model and the observational constraints, is the HXLF at  
$L_X$ $\lesssim$ 10$^{42}$ erg s$^{-1}$, specifically at $z$ = 0.6.  Due to the tendril nature of cosmic structure, narrow fields, 
which tend to provide the high $z$, low $L_X$ AGN detections, will observe over dense regions when one of 
these tendrils passes through the observed field.  This cosmic variance may cause the number counts and HXLFs observed in 
different fields to vary slightly \citep{BH05}.  Observations of AGN with $L_X$ $\lesssim$ 10$^{42}$ erg s$^{-1}$ and $z$ $\sim$ 
0.6 can only be conducted in the {\em Chandra} deep fields, which are known to have significant field-to-field variation \citep{Cowie02}.
As the slope of the low $L_X$ end of the HXLF is still fairly uncertain, 
cosmic variance is most likely to affect the low $L_X$ end of the observed HXLF.

When both merger triggered and secularly triggered AGN are considered, models which use the \citet{N09} ABHMF are in 
agreement with observations at $z$ $<$ 0.5, but not at high redshift.  In contrast, models which use 
the \citet{MH08} ABHMF for both 
merger and secularly triggered AGN over-predict the black hole mass density.  In 
order to explain both high and low redshift observations, it is necessary that merger triggered and secularly triggered AGN 
have different ABHMFs which evolve differently with redshift.  The mass distribution of black holes triggered by secular 
processes is, therefore, intrinsically different than 
the mass distribution of black holes triggered by mergers, in agreement with the findings of \citet{S10}.




\section{Discussion}
\label{sect:disc}

An observationally based AGN triggering rate and a theoretical AGN light curve is used to model the AGN population.
By considering the HXLF at five different redshifts, the black hole mass density, the XRB spectrum, and the AGN number 
counts in the 2--10 and 15--55 keV bands, it is determined that neither merger triggered AGN nor AGN triggered by 
secular processes can alone account for the entire AGN population.  Two populations are necessary to describe the 
observed evolution and space density of AGN.  Furthermore, it is found that the ABHMF of merger 
triggered AGN evolves more rapidly and has a higher average black hole mass than the ABHMF of secularly triggered AGN.

\subsection{AGN Light Curve Model}
\label{sub:lcm}

The AGN light curve model used has three parameters, $t_0$, $\beta$, and $\eta$.  The $\eta$ parameter is related to the peak Eddington 
ratio of the light curve.  The AGN population model is least sensitive to $\eta$, however, it is found that $\eta$ $\gtrsim$ 1.0 provides 
the best fit to the observational constraints, suggesting that short periods of super-Eddington accretion are likely in a significant 
fraction of the AGN population.  The AGN population model is moderately dependent on the time scale parameter, $t_0$.  If this time scale 
is too short, the AGN fade too quickly, and the space density of low luminosity AGN is severely under-predicted.  Thus, $t_0$ is most 
important for the normalization of the AGN HXLF.  The shape of the HXLF is primarily controlled by the slope of the light 
curve, which is controlled by the $\beta$ parameter.  The AGN population model is most sensitive to $\beta$.  Changing $\beta$ by 0.1 can cause 
the resulting HXLF shape to change substantially.  The primary goal of this study is to investigate if it is necessary that AGN be triggered by 
both mergers and secular processes or if a single mechanism can account for the entire AGN population, not to determine the best fit AGN light curve 
parameters.  However, the observational constraints used in this study do provide interesting restrictions on the AGN light curve parameters.  

\citet{HH09} point out that different models of the AGN light curve predict different light curve slopes, $\beta$.  Therefore, 
the AGN light curve fit provides information about the physical conditions in the vicinity of the accretion disk.  For example, 
self-regulated models, where the black hole accretion is feedback limited, predict $\beta$ = 0.3 -- 0.8.  Meanwhile, if AGN are 
fueled by mass loss from a nuclear star cluster, $\beta$ = 0.9 -- 1.0.  Models in which the fuel supply of a \citet{SS73} 
accretion disk is suddenly cut off, predict $\beta$ = 0.80 -- 0.84.  In all of the best fit merger triggered AGN models presented here, 
$\beta$ = 0.7, suggesting that for merger triggered AGN, the accretion is feedback limited.  Also, for the best fit model presented here, 
$\eta$, the parameter related to light curve peak Eddington ratio, is much smaller for AGN triggered by secular mechanisms than for AGN triggered by 
major mergers.

\subsection{Two Populations of AGN}

\begin{figure*}
\begin{center}
\includegraphics[angle=0,width=0.95\textwidth]{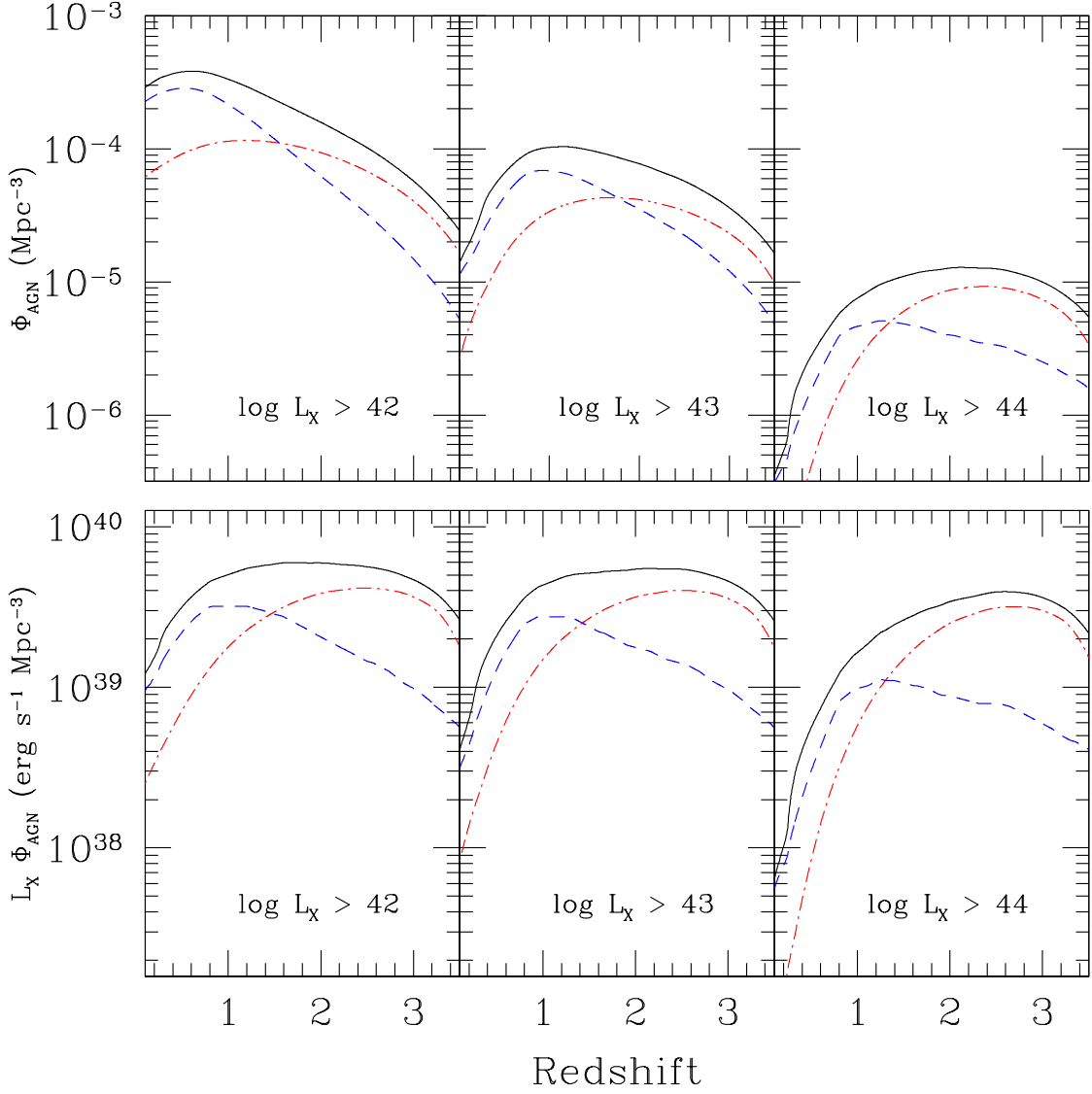}
\end{center}
\caption{The space and luminosity density of AGN as a function of redshift predicted by the mixed ABHMF model.  The top row shows the space density of AGN for $\log L_X$ $>$ 42, 43, and 44.  The bottom row shows the luminosity density of AGN for $\log L_X$ $>$ 42, 43, and 44.  The line styles are the same as in Figure \ref{fig:mixedlf}.}
\label{fig:den}
\end{figure*}

Here, it is assumed that every major merger triggers an AGN and major mergers can account for 
the AGN population at $z$ = 2.3, thus this analysis is an upper limit for the contribution of major merger triggered AGN.  
Still, it is found that at $z$ $\lesssim$ 2, a significant fraction 
of all AGN are triggered through secular mechanisms.  Figure \ref{fig:den} shows the space and luminosity density of AGN with 
$\log L_X$ $>$ 42, 43, and 44, which are triggered by mergers and secular processes as a function of redshift for the mixed ABHMF model.  
At all three luminosity ranges, the luminosity density and space density of AGN is dominated by major merger triggered AGN at $z$ $\gtrsim$ 
1.5 and secularly triggered AGN at $z$ $\lesssim$ 1.5.  By varying model parameters, we find the minimum 
redshift at which major mergers can dominate AGN trigger is $z$ $\sim$ 1.  This finding is in agreement with \citet{G09}, who found that $\sim$30$\%$ of the AGN space density 
and $\sim$25$\%$ of the AGN luminosity density at $z$ $\sim$ 1 is due to AGN hosted by disk dominated hosts, implying that secular 
processes are responsible for at least one quarter of the AGN luminosity density at $z$ $\sim$ 1.  It is found that merger triggered 
AGN account for $\sim$20$\%$ of the space density of $\log L_X$ $>$ 43 AGN at $z$ $<$ 0.05, consistent with findings of \citet{K10}.  
Thus, AGN hosted by disk dominated galaxies are a significant fraction of the AGN population by $z$ $\sim$ 1, indicating that secular 
evolution is an important mode of galaxy evolution at this redshift.

Similarly, \citet{DB11b} found that for AGN at $z$ $\lesssim$ 1 the host galaxies of obscured 
and unobscured AGN are, on average, the same, suggesting that at $z$ $\lesssim$ 1 AGN and galaxy evolution is controlled by secular 
processes.  However, at $z$ $\gtrsim$ 1, the host galaxies of unobscured AGN are intrinsically 
less dusty than the host galaxies of obscured AGN at similar redshifts, suggesting a fundamental change in 
the mechanisms which control AGN activity at $z$ $\sim$ 1 \citep{DB11b}. Indeed, Figure \ref{fig:mixedlf} shows that, at $z$ $\lesssim$ 1, 
AGN triggered through secular processes dominate even the high $L_X$ end of the HXLF, while fading major merger triggered AGN dominate the 
low $L_X$ end of the HXLF.  These findings are also in agreement with the conclusions of a variety of recent 
observational studies which found that secular processes are an important form, and possibly 
the dominate form, of galaxy evolution at $z$ $\lesssim$ 2 \citep{G09,A11,C11,S11,W11}.  Indeed, according to the HXLF, 
we find that AGN triggered by secular processes can account for the entire AGN population with $\log L_X$ $>$ 43 at $z$ $\lesssim$ 0.5.  
\citet{HH06} suggest that secularly triggered  AGN are not cosmologically important; however, the findings of this study illustrate that AGN triggered by 
secular mechanisms are necessary to describe the AGN population and dominate the space density of AGN with $\log L_X$ $>$ 43 at $z$ $\lesssim$ 1.5.

The findings of \citet{S10} suggest that this fundamental change in AGN activity at $z$ $\sim$ 1.5 is due to cosmic downsizing.  
By studying AGN black hole masses as a function of host galaxy morphology for AGN observed by the Sloan Digital 
Sky Survey (SDSS),  
\citet{S10} find that it is preferentially the least massive black holes in less massive early type galaxies (stellar 
mass $M_*$ $\sim$ 10$^{10}$ M$_{\odot}$) which are currently active.  In contrast, the black holes which are currently 
active in late type galaxies are preferentially the most massive black holes hosted by more massive late type host 
galaxies ($M_*$ $\sim$ 10$^{11}$ M$_{\odot}$).  A stellar disk generally indicates that a galaxy has not undergone a 
major merger or that the last major merger experienced by the galaxy was long enough ago that a disk had time to reform, 
$\gtrsim$ 500 Myr \citep{Ho11}.  We can therefore assume that, for AGN hosted by late type galaxies, the current episode 
of AGN activity was triggered by secular processes.  \citet{S10} suggest that the early type galaxy hosts in their sample 
may be smaller versions of the mode of galaxy evolution experienced by massive ellipticals at high redshift.  Thus, the 
early type host galaxies in the sample of \citet{S10} may host downsized versions of major merger triggered AGN.
Thus, it appears that the dominance of major mergers in galaxy evolution began to decline by $z$ $\sim$ 2 and 
is continuing to decline, in agreement with the theory of cosmic downsizing.  Meanwhile, the importance of secular evolution 
increased as the importance of major mergers decreased, until secular evolution became the dominate form of galaxy evolution 
at $z$ $\sim$ 1.5.  Since we are still early in the era of secularly dominated galaxy evolution, it is the more massive systems 
which are currently undergoing secularly triggered AGN activity.  Cosmic downsizing would therefore predict that as the era of 
secular dominance continues, AGN triggered through secular processes will become more common in lower mass systems. 


\subsection{ABHMF and Evolution}

It is found that AGN triggered by major mergers at $z$ $<$ 4 have, on average, more massive central black 
holes than AGN triggered by secular processes, at least at high redshift.  Locally, it appears that merger triggered AGN and secularly 
triggered AGN have similar black holes masses \citep{S10}, necessitating that the ABHMF of merger triggered AGN 
evolves differently than the ABHMF of secularly triggered AGN.  The ABHMF of AGN triggered by secular mechanisms appears to 
evolve in a manner consistent with the continuity equation (Eq. \ref{eq:coneq}), which describes the evolution of the combined 
active and quiescent black hole mass function \citep[e.g.,][]{SB92, MH08}.  However, if the ABHMF of merger triggered AGN evolves following 
the continuity equation, then at $z$ $\gtrsim$ 1 the average active black hole mass is too small to account for the high 
$L_X$ end of the HXLF.  In contrast, if the \citet{L09} ABHMF evolution is used to evolve the ABHMF of secularly triggered AGN, 
then the low $L_X$ end of the $z$ $\lesssim$ 1 HXLF is significantly under-predicted.  Thus, it appears that not only are both 
secular processes and major mergers important mechanisms for triggering the $z$ $<$ 2 AGN population, but these two mechanisms 
trigger different populations of black holes and these two populations have different redshift evolution.

It is not surprising that the ABHMF of merger triggered AGN evolves differently from the secularly triggered ABHMF and the 
quiescent black hole mass function.  The major merger rate evolves strongly with redshift and galaxy mass \citep{H10}.  Thus, it is 
expected that the ABHMF of AGN triggered by major mergers would also evolve strongly with redshift \citep[e.g.,][]{T11}.  In contrast, 
the rate of AGN being triggered by secular evolution, $f_{sec}$, is assumed to be constant with redshift.  Therefore, it is expected 
that the ABHMF for AGN triggered by secular processes will be less redshift dependent than the ABHMF for merger triggered AGN.
Also, as the majority of supermassive black holes at any redshift are quiescent, the combined quiescent and active black hole mass function 
evolves more slowly with redshift than the merger triggered ABHMF.  Thus, it is expected that the merger triggered ABHMF will evolve more 
strongly with redshift than the ABHMF of secularly triggered AGN or of the combined active and quiescent black hole mass function.

\subsection{The Connection Between Compton Thick AGN and Mergers}
\label{sub:ct}

Observational and theoretical evidence suggests that Compton thick AGN are AGN which were recently triggered by a mechanism which has 
caused a large amount of gas and dust to be funneled into the nuclear regions of the host galaxy \citep[e.g.,][]{F99, P04, F08, F09, DB10, T10, DB11b}.  
Due to the large amount of gas and dust required for the rapid AGN fueling and high column density necessary for an AGN to be Compton 
thick, the most likely triggering process is a wet major merger \citep{S88,H06}, though \citet{Wein11} argue that Compton thick levels of 
obscuration are also possible in instances of clumpy cold-flow accretion onto the host galaxy.  Here, we set $f_{CT}$ = 0.0 for the secularly 
triggered AGN and calculate $f_{CT}$ for the merger triggered AGN as described in Section \ref{sub:HXLF}.  If it is assumed that Compton thickness 
is an evolutionary phase which only merger triggered AGN experience, the $\chi_{red}^2$ of the mixed ABHMF increases by 0.1 and the model is 
still in agreement with the observed black hole space density, XRB spectrum, and the 2--10 keV and 15--55 keV number counts.  Therefore, the hypothesis that Compton thick AGN are 
an evolutionary stage of merger triggered AGN is fully consistent with the model presented here.

\section{Summary}
\label{sect:sum}

By combining an observationally motivated AGN triggering rate and a theoretically motivated AGN light curve, an AGN population synthesis model 
is developed which can track the evolution of AGN triggered by mergers and secular processes.  The light curve parameters are 
constrained by comparing model predictions against the observed AGN HXLF.  The observed space density of supermassive black holes, 
AGN number counts, and XRB spectrum are also considered.  The results of this study are summarized below.

$\bullet$ Neither AGN triggered by major mergers nor AGN triggered by secular mechanisms alone are capable of accounting for the entire observed AGN population at $z$ $\lesssim$ 2.  This finding is independent of the SMF used or the potential AGN host galaxy minimum mass.


$\bullet$ The dominant AGN triggering mechanism switches from major mergers at $z$ $\gtrsim$ 1.5 to secular mechanisms by $z$ $\sim$ 1.  This is a natural consequence of both the major merger rate and fraction of gas rich galaxies increasing with redshift.  However, both major mergers and secular processes are important triggering mechanisms for AGN over the entire redshift range considered, especially at moderate luminosities.  .

$\bullet$ The black hole mass density is dominated by black hole growth triggered by major mergers over the entire redshift range considered.  

$\bullet$ The space density and luminosity density of AGN with $\log L_X$ $>$ 42, 43, and 44 is dominated by AGN triggered by major mergers at $z$ $\gtrsim$ 1.5 and by AGN triggered by secular processes at $z$ $\lesssim$ 1.5.  

$\bullet$ The XRB is dominated by emission from AGN triggered by secular mechanisms.

$\bullet$ The ABHMF of AGN triggered by major mergers is different from the ABHMF of AGN triggered by secular mechanisms.  At high redshift, the average black hole triggered by major mergers is more massive than the average black hole triggered by secular processes, which is consistent with cosmic downsizing.

$\bullet$ The evolution of the ABHMF of AGN triggered by major mergers is also different from the evolution of the ABHMF of AGN triggered through secular mechanisms, with the ABHMF of major merger triggered AGN evolving more rapidly with redshift than the ABHMF of AGN triggered by secular processes.  The observed evolution of \citet{L09} appears to describe the evolution of the merger triggered AGN ABHMF well.  In contrast, the continuity equation (Eq. \ref{eq:coneq}) describes the evolution of the ABHMF for secularly triggered AGN.  This finding is consistent with the strong redshift evolution of the major merger rate.

$\bullet$ The findings of this study are not changed if Compton thick AGN are an evolutionary stage only experienced by major merger triggered AGN.

\acknowledgments
This work was supported by NSF award AST 1008067.  The authors thank the referee for helpful comments which improved this paper.


{}

%
\begin{deluxetable}{lccc|cccc}
\tabletypesize{\small}
\tablecolumns{8}
\tablecaption{Summary of derived model free parameters.}
\tablehead{
  \colhead{Model} &
  \multicolumn{3}{c} {Merger Triggered AGN} &
  \multicolumn{4}{c} {Secularly Triggered AGN} \\
  \colhead{} &
  \colhead{$t_0$ (yr)} &
  \colhead{$\beta$} &
  \colhead{$\eta$} &
  \colhead{$t_0$ (yr)} &
  \colhead{$\beta$} &
  \colhead{$\eta$} &
  \colhead{$f_{sec}$ (Gyr$^{-1}$)} \\
}

\startdata
  \cutinhead{One Population Models}
  Mergers only & 2.5 $\times$ 10$^8$ & 0.7 & 2.5 & - & - & - & 0.0 \\
  Secular only & - & - & - & 2.5 $\times$ 10$^8$ & 0.8 & 0.4 & 0.02 \\
  \cutinhead{Two Population Models}
  N09 w/ Eq. \ref{eq:coneq} & 2.5 $\times$ 10$^8$ & 0.7 & 0.4 & 5 $\times$ 10$^8$ & 0.5 & 2.5 & 0.005 \\
  N09 w/ L09 & 1 $\times$ 10$^8$ & 0.5 & 1.0 & 2.5 $\times$ 10$^8$ & 0.6 & 2.5 & 0.005 \\
  MH08 w/ Eq. \ref{eq:coneq} & 2.5 $\times$ 10$^8$ & 0.8 & 2.5 & 5 $\times$ 10$^8$ & 0.7 & 2.5 & 0.01 \\
  MH08 w/ L09 & 2.5 $\times$ 10$^8$ & 0.7 & 2.5 & 5 $\times$ 10$^8$ & 0.8 & 2.5 & 0.005 \\
  \cutinhead{Two Population Mixed ABHMF Model}
  Mixed ABHMF & 2.5 $\times$ 10$^8$ & 0.7 & 2.5 & 2.5 $\times$ 10$^8$ & 0.8 & 2.5 & 0.01 \\
\enddata

\tablecomments{MH08 refers to the \citet{MH08} ABHMF, N09 refers to the \citet{N09} ABHMF, and L09 refers to the \citet{L09} ABHMF evolution.  The mixed ABHMF model uses the MH08 ABHMF with L09 evolution for the merger triggered AGN and the N09 ABHMF with the continuity equation (Eq. \ref{eq:coneq}) evolution for the secularly triggered AGN.}
\label{modelparam}
\end{deluxetable}
\begin{deluxetable}{lcccccc}
\tabletypesize{\small}
\tablecolumns{6}
\tablecaption{Summary of model fits to the observational constraints.}
\tablehead{
  \colhead{Model} &
  \colhead{d.o.f.\tablenotemark{a}} &
  \colhead{$\chi^2_{red}$\tablenotemark{b}} &
  \colhead{$\rho_{\bullet}$\tablenotemark{c}} &
  \colhead{XRB\tablenotemark{d}} &
  \colhead{2--10 keV\tablenotemark{e}} & 
  \colhead{15--55 keV\tablenotemark{f}} \\
}
\startdata
  \cutinhead{One Population Models}
  Mergers only & 123 & 2.4 & $\sqrt{}$ & $\times$  & $\times$ & $\times$ \\
  Secular only & 122 & 2.1 & $\times$  & $\times$  & $\times$ & $\times$  \\
  \cutinhead{Two Population Models}
  N09 w/ Eq. \ref{eq:coneq} & 119 & 1.5 & $\times$ & $\times$ & $\sqrt{}$ & $\sqrt{}$ \\
  N09 w/ L09 & 119 & 1.4 & $\times$  & $\times$ & $\sqrt{}$ & $\sqrt{}$ \\
  MH08 w/ Eq. \ref{eq:coneq} & 119 & 1.7 & $\times$ & $\times$ & $\times$ & $\times$ \\
  MH08 w/ L09 & 119 & 1.5 & $\sqrt{}$ & $\times$ & $\times$ & $\times$ \\
  \cutinhead{Two Population Mixed ABHMF Model}
  Mixed ABHMF\tablenotemark{g} & 119 & 1.4 & $\sqrt{}$ & $\sqrt{}$ & $\sqrt{}$ & $\sqrt{}$ \\
\enddata

\tablenotetext{a}{The number of degrees of freedom for the $\chi^2$ test reported in column $\chi^2_{red}$.}
\tablenotetext{b}{The quoted $\chi^2_{red}$ refers to the 127 HXLF data points plotted in Figure \ref{fig:one}.}
\tablenotetext{c}{A $\times$ in this column denotes that, when considering the two black hole space density data points, $\chi^2$ $>$ 4.0 for the designated model.}
\tablenotetext{d}{A $\times$ in this column denotes that this model either over or under-predicts the XRB by more than 25$\%$ in the 10--50 keV band.}
\tablenotetext{e}{A $\times$ in this column denotes that, this model either over or under-predicts the the 2--10 keV observed number counts shown in Figure \ref{fig:2to8} by more than 25$\%$ at all fluxes.}
\tablenotetext{f}{A $\times$ in this column denotes that, this model either over or under-predicts the 15--55 keV observed $\log N$--$\log S$ shown in Figure \ref{fig:bat} by more than 25$\%$ at all fluxes.}
\tablenotetext{g}{The mixed ABHMF model uses the MH08 ABHMF with L09 evolution for the merger triggered AGN and the N09 ABHMF with the continuity equation (Eq. \ref{eq:coneq}) evolution for the secularly triggered AGN.}

\label{modelfit}
\end{deluxetable}

\end{document}